\documentclass[useAMS,usenatbib]{mn2e}

\usepackage{graphicx}
\usepackage{epsfig,floatflt}
\usepackage{times}

\title[New spectroscopic binaries among Galactic Cepheids]
{Discovery of the spectroscopic binary nature 
of six southern Cepheids}

\author[Szabados et al.]{L. Szabados$^{1}$,  A. Derekas$^{1,2}$, 
L. L. Kiss$^{1,2,3}$, J. Kov\'acs$^3$, R.~I. Anderson$^4$, Cs. Kiss$^1$, 
\newauthor T. Szalai$^5$, P. Sz\'ekely$^6$, J.~L. Christiansen$^7$
\\$^1$Konkoly Observatory, Research Centre for Astronomy and Earth 
Sciences, Hungarian Academy of Sciences, H-1121 Budapest, \\ 
Konkoly Thege Mikl\'os \'ut 15-17, Hungary\\
$^2$Sydney Institute for Astronomy, School of Physics, 
University of Sydney, NSW 2006, Australia\\
$^3$ELTE Gothard-Lend\"ulet Research Group, H-9700 Szombathely, 
Szent Imre herceg \'ut 112, Hungary\\
$^4$Observatoire de Gen\`eve, Universit\'e de Gen\`eve, 
51 Ch. des Maillettes, CH-1290 Versoix, Switzerland\\
$^5$Department of Optics and Quantum Electronics, 
University of Szeged, D\'om t\'er 9, H-6720 Szeged, Hungary\\
$^6$Department of Experimental Physics, University of Szeged, 
Szeged H-6720, Hungary\\
$^7$SETI Institute/NASA Ames Research Center, M/S 244-30, 
Moffett Field, CA 94035, USA
}

\begin{document}

\date{Accepted Received ; in original form }

\pagerange{\pageref{firstpage}--\pageref{lastpage}} \pubyear{2012}

\maketitle

\label{firstpage}

\begin{abstract}
We present the analysis of photometric and spectroscopic data 
of six bright Galactic Cepheids: GH~Carinae, V419~Centauri, 
V898~Centauri, AD~Puppis, AY~Sagittarii, and ST~Velorum. 
Based on new radial velocity data (in some cases supplemented 
with earlier data available in the literature), these Cepheids 
have been found to be members in spectroscopic binary systems. 
V898~Cen turned out to have one of the largest orbital radial 
velocity amplitude ($> 40$ km\,s$^{-1}$) among the known binary
Cepheids. The data are insufficient to determine the orbital 
periods nor other orbital elements for these new spectroscopic 
binaries.

These discoveries corroborate the statement on the high frequency 
of occurrence of binaries among the classical Cepheids, a fact 
to be taken into account when calibrating the period-luminosity 
relationship for Cepheids. 

We have also compiled all available photometric data 
that revealed that the pulsation period of AD~Pup, the longest 
period Cepheid in this sample, is continuously increasing with 
$\Delta P={\rm 0.004567 d/century}$, likely to be caused by stellar 
evolution. The wave-like pattern superimposed on the parabolic 
$O-C$ graph of AD~Pup may well be caused by the light-time effect 
in the binary system. ST~Vel also pulsates with a continuously 
increasing period. The other four Cepheids are characterised 
with stable pulsation periods in the last half century.

\end{abstract}


\section{Introduction}
\label{intro}

Classical Cepheid variable stars are primary distance
indicators because owing to the famous period-luminosity 
($P$-$L$) relationship they rank among standard candles 
in establishing the cosmic distance scale.

Companions to Cepheids, however, complicate the situation.
The contribution of the secondary star to the observed 
brightness has to be taken into account when involving any
particular Cepheid in the calibration of the $P$-$L$ relationship.
Binaries among Cepheids are not rare at all: their frequency 
exceeds 50 per cent for the brightest Cepheids, while among the 
fainter Cepheids an observational selection effect encumbers 
revealing binarity \citep{Sz03a}.

It is essential to study Cepheids individually from the point 
of view of binarity before involving them in any calibration 
procedure (of e.g. $P$-$L$ or period-radius relationship).
This attitude is especially important if Cepheid-related
relationships are calibrated using a small sample. However, a
deep observational analysis of individual Cepheids can only be 
performed in the case of their Galactic representatives. When
dealing with extragalactic Cepheids, unrevealed binarity is one
of the factors that contribute to the dispersion of the 
$P$-$L$ relationship. A detailed list of physical factors 
responsible for the finite width of the $P$-$L$ relationship 
around the ridge line approximation is given by \citet{SzK12}.

The orbital period of binaries involving a supergiant Cepheid 
component cannot be shorter than about a year. 
Spectroscopic binaries involving a Cepheid component with orbital 
periods longer than a decade are also known (see the on-line data 
base on binaries among Galactic Cepheids:
{\tt http://www.konkoly.hu/CEP/orbit.html}).
Therefore, a first-epoch radial velocity curve, especially based
on data obtained in a single observational season, is usually
insufficient for pointing out an orbital effect superimposed
on the radial velocity changes due to pulsation.

In the case of pulsating variables, like Cepheids, spectroscopic
binarity (SB) manifests itself in a periodic variation of the
$\gamma$-velocity (i.e., the radial velocity of the mass centre
of the Cepheid). In practice, the orbital radial velocity variation
of the Cepheid component is superimposed on the radial velocity
variations of pulsational origin. To separate the orbital and 
pulsational effects, knowledge of the accurate pulsation period
is essential, especially when comparing radial velocity
data obtained at widely differing epochs. Therefore, the pulsation
period and its variations have been determined with the method of
the $O-C$ diagram \citep{S05} for each target Cepheid. Use of the
accurate pulsation period obtained from the photometric data is a 
guarantee for the correct phase matching of the (usually less
precise) radial velocity data.

In this paper we point out spectroscopic binarity of six 
bright Galactic Cepheids by analysing radial velocity data.
The structure of this paper is as follows. The new observations
and the equipments utilized are described in Section~\ref{newdata}.
Section~\ref{results} are devoted to the results on the six new 
SB Cepheids: GH~Carinae, V419~Centauri, V898~Centauri, AD~Puppis, 
AY~Sagittarii, and ST~Velorum, respectively. Basic information on 
these Cepheids are found in Table~\ref{obsprop}. Finally, 
Section~\ref{concl} contains our conclusions.

\begin{table}  
\begin{center}  
\caption{Basic data of the programme stars 
and the number of spectra} 
\label{obsprop}  
\begin{tabular}{|rcccc|} 
\hline  
Star &  $\langle V \rangle$ & P & Mode & No. of obs.\\
& (m) & (d)& of pulsation &\\
\hline  
GH~Car   &  9.18 &  5.725532 & first overtone & 27+43\\
V419~Cen &  8.19 &  5.507123 & first overtone & 26\\ 
V898~Cen &  8.00 &  3.527310 & first overtone & 33+4\\
AD~Pup   &  9.91 & 13.596919 & fundamental   & 33\\  
AY~Sgr   & 10.55 &  6.569667 & fundamental   & 22\\ 
ST~Vel   &  9.73 &  5.858316 & fundamental   & 27\\   
\hline   
\end{tabular} 
\end{center}  
\end{table}


\section{New observations}
\label{newdata}

\subsection{Spectra from Siding Spring Observatory}
\label{SSO}

We performed a radial velocity (RV) survey of Cepheids with 
the 2.3~m ANU telescope located at Siding Spring Observatory, 
Australia. The main aim of the project was to detect Cepheids 
in binary systems by measuring changes in the mean values of 
their RV curve which can be interpreted as the 
orbital motion of the Cepheid around the center of mass in 
a binary system (change of $\gamma$-velocity). 
The target list was compiled to include Cepheids with
single-epoch radial velocity phase curve or without any
published radial velocity data. Several Cepheids suspected 
members in spectroscopic binaries were also put on the
target list. On 64 nights between October 2004 and March 2006 
we monitored 40 Cepheids with pulsation periods between 2 and 30 
days. (V898~Cen was observed in July 2009, too).

Medium-resolution spectra were taken with the Double Beam 
Spectrograph using the 1200~mm$^{-1}$ gratings in both arms of 
the spectrograph. The projected slit width was 2$^{\prime\prime}$ 
on the sky, which was about the median seeing during our 
observations. The spectra covered the wavelength ranges 
4200--5200~\AA\ in the blue arm and 5700--6700~\AA\ in the red 
arm. The dispersion was 0.55~\AA~px$^{-1}$, leading to a nominal 
resolution of about 1~\AA.

All spectra were reduced with standard tasks in {\sc iraf}
\footnote{{\sc iraf} is distributed by the National Optical 
Astronomy Observatories, which are operated by the Association of 
Universities for Research in Astronomy, Inc., under cooperative 
agreement with the National Science Foundation.}.
Reduction consisted of bias and flat field corrections, 
aperture extraction, wavelength calibration, and continuum 
normalisation. We checked the consistency of wavelength 
calibrations via the constant positions of strong telluric 
features, which proved the stability of the system. 
Radial velocities were determined only for the red arm data 
with the task {\it fxcor\/}, applying the cross-correlation 
method using a well-matching theoretical template spectrum 
from the extensive spectral library of \citet{Metal05}. Then, we 
made barycentric corrections to every single RV 
value. This method resulted in a 1-2~km~s$^{-1}$ uncertainty 
in the individual radial velocities, while further tests have 
shown that our absolute velocity frame was stable to within 
$\pm$2-3~km~s$^{-1}$. This level of precision is sufficient
to detect a number of Cepheid companions, as they can often
cause $\gamma$-velocity changes well above 10~km~s$^{-1}$.

\subsection{FEROS observations in ESO}
\label{feros}

V898~Centauri was observed on four consecutive nights in April, 
2011, using the \textit{FEROS} (Fiber-fed Extended Range Optical 
Spectrograph) instrument on the MPG/ESO 2.2\,m telescope 
in La Silla Observatory, Chile (see Table~\ref{tab-v898cen-feros-data}).
The \textit{FEROS} has a total wavelength coverage of 356-920\,nm 
with a resolving power of $R=48\,000$  \citep{FEROS1,FEROS2}.
Two fibres, with entrance aperture of 2\farcs7, simultaneously
recorded star light and sky background. The detector is a 
back-illuminated CCD with 2948$\times$4096 pixels of 15\,$\mu$m
size. Basic data reduction was performed using a pipeline package 
for reductions (DRS), in {\sc midas} environment. The pipeline 
performs the subtraction of bias and scattered light in the CCD, 
orders extraction, flat-fielding and wavelength calibration with 
a ThAr calibration frame (the calibration measurements were 
performed at the beginning of each night, using the ThAr lamp).

After the continuum normalization of the spectra using {\sc iraf}
we determined the radial velocities with the task {\it fxcor\/}, 
as in the case of the SSO spectra (see Sect.~\ref{SSO}). The 
velocities were determined in the region 500-600~nm where a number 
of metallic lines are present and hydrogen lines are lacking.
We made barycentric corrections to each RV value 
with the task {\it rvcorrect\/}. The estimated uncertainty of the 
radial velocities is 0.05~km\,s$^{-1}$.

\subsection{CORALIE observations from La Silla}
\label{coralie}

GH~Car was among the targets during multiple observing 
campaigns between April 2011 and May 2012 using the fiber-fed 
high-resolution ($R \sim 60000$) echelle spectrograph 
\textit{CORALIE} mounted at the Swiss 1.2\,m Euler telescope at 
ESO La Silla Observatory, Chile. The instrument's design is 
described in \citet{Qetal01}; recent instrumental updates 
are found in \citet{Wetal08}. 

The spectra are reduced by the efficient online reduction 
pipeline that performs bias correction, cosmics removal, 
and flatfielding using tungsten lamps. ThAr lamps are used 
for the wavelength calibration. The reduction pipeline directly 
determines the RV through cross-correlation \citep{Betal96} 
using a mask that resembles a G2 spectral type. 
The RV stability of the instrument is excellent and for 
non-pulsating stars the RV precision is limited by photon-noise, 
see e.g., \citet{Petal02}. However, the precision achieved for 
Cepheids is lower due to line asymmetries. We estimate a typical 
precision of $\sim$ 0.1\,km\,s$^{-1}$ (including systematics due
to pulsation) per data point for our data. The radial velocity 
data are listed in Table~\ref{tab-ghcar-coralie-data}.

\section{Results}
\label{results}

\subsection{GH Carinae}
\label{ghcar}

\paragraph*{Accurate value of the pulsation period}
\label{ghcar-period}

The brightness variability of GH~Car (HD\,306077, $\langle V \rangle
= 9.18$\,mag.) was revealed by Oosterhoff \citep{H30}. This
Cepheid pulsates in the first overtone mode, therefore it has a small
pulsational amplitude and nearly sinusoidal light (and velocity)
curve. 

In the case of Cepheids pulsating with a low amplitude, 
the $O-C$ diagram constructed for the median brightness is more 
reliable than that based on the moments of photometric maxima 
\citep{Detal12}. Therefore we determined the accurate value
of the pulsation period by constructing an $O-C$ diagram for
the moments of median brightness (the mid-point between the
faintest and the brightest states) on the ascending branch of 
light curve since it is this phase when the brightness variations 
are steepest during the whole pulsational cycle.

All published observations of GH~Car covering half a century were 
re-analysed in a homogeneous manner to determine seasonal moments 
of the chosen light curve feature. The relevant data listed in 
Table~\ref{tab-ghcar-oc} are as follows:\\
Column~1: Heliocentric moment of the median brightness 
on the ascending branch;\\
Col.~2: Epoch number, $E$, as calculated from Eq.~1:
\vspace{-1mm}
\begin{equation}
C = 2\,451\,266.2198 + 5.725\,532{\times}E 
\label{ghcar-ephemeris}
\end{equation}
\vspace{-3mm}
$\phantom{mmmmm}\pm0.0038\phantom{}\pm 0.000\,004$

\noindent (this ephemeris has been obtained by the weighted least 
squares fit to the tabulated $O-C$ differences);\\
\noindent Col.~3: the corresponding $O-C$ value as calculated 
from Eq.~\ref{ghcar-ephemeris};\\
Col.~4: weight assigned to the $O-C$ value (1, 2, or 3 
depending on the quality of the light curve leading to 
the given difference);\\
Col.~5: reference to the origin of data, preceded by the name 
of the observer if different from the author(s) cited.\\

\begin{figure}
\includegraphics[height=44mm, angle=0]{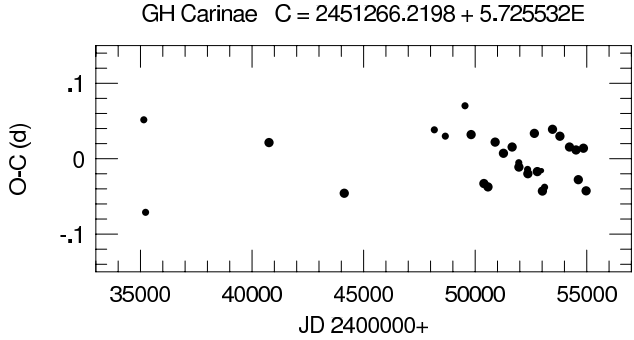}
\caption{$O-C$ diagram of GH~Car based on the values
listed in Table~\ref{tab-ghcar-oc}. The pulsation period
of GH~Car is constant}
\label{fig-ghcar-oc}
\end{figure}

The plot of $O-C$ values shown in Fig.~\ref{fig-ghcar-oc}
can be approximated with a constant period. The scatter of the 
points in the figure reflects the observational error and 
uncertainties in the analysis of the data.

\begin{table}
\caption{$O-C$ values of GH~Carinae (see the description 
in Sect.~\ref{ghcar-period})}
\begin{tabular}{l@{\hskip2mm}r@{\hskip2mm}r@{\hskip2mm}c@{\hskip2mm}l}
\hline
\noalign{\vskip 0.2mm}
JD$_{\odot}$ & $E\ $ & $O-C$ & $W$ & Data source\\
2\,400\,000 + &&&\\
\noalign{\vskip 0.2mm}
\hline
\noalign{\vskip 0.2mm}
35148.8988&$-$2815 &+0.0516 &2& \citet{Wetal58}\\
35228.9336&$-$2801&$-$0.0711 &2& \citet{I61}\\
40759.8900&$-$1835 &+0.0214 &3& \citet{P76}\\
44132.1611&$-$1246&$-$0.0458 &3& \citet{B08}\\
48168.7453 &$-$541 &+0.0383 &2& {\it Hipparcos} \citep{ESA97}\\
48661.1326 &$-$455 &+0.0299 &2& {\it Hipparcos} \citep{ESA97}\\
49542.9049 &$-$301 &+0.0702 &2& \citet{B08}\\
49817.6922 &$-$253 &+0.0320 &3& \citet{B08}\\
50390.1805 &$-$153&$-$0.0329 &3& \citet{B08}\\
50573.3930 &$-$121&$-$0.0374 &3& \citet{B08}\\
50894.0822 &$-$65 &+0.0220 &3& \citet{B08}\\
51266.2270 & 0 &+0.0072 &3& \citet{B08}\\
51655.5715 &    68 &+0.0155 &3& \citet{B08}\\
51953.2783 &   120&$-$0.0053 &2& ASAS \citep{P02}\\
51958.9981 &   121&$-$0.0111 &3& \citet{B08}\\
52348.3311 &   189&$-$0.0142 &2& ASAS \citep{P02}\\
52359.7765 &   191&$-$0.0199 &3& \citet{B08}\\
52651.8321 &   242 &+0.0336 &3& \citet{B08}\\
52783.4687 &   265&$-$0.0171 &3& ASAS \citep{P02}\\
52972.4125 &   298&$-$0.0158 &1&  {\it INTEGRAL} OMC\\
53012.4642 &   305&$-$0.0429 &3& \citet{B08}\\
53109.8033 &   322&$-$0.0378 &2& ASAS \citep{P02}\\
53464.8631 &   384 &+0.0390 &3& ASAS \citep{P02}\\
53796.9347 &   442 &+0.0298 &3& ASAS \citep{P02}\\
54226.3353 &   517 &+0.0155 &3& ASAS \citep{P02}\\
54518.3336 &   568 &+0.0116 &3& ASAS \citep{P02}\\  
54621.3538 &   586&$-$0.0278 &3& ASAS \citep{P02}\\
54850.4167 &   626 &+0.0139 &3& ASAS \citep{P02}\\
54970.5964 &   647&$-$0.0426 &3& ASAS \citep{P02}\\
\noalign{\vskip 0.2mm}
\hline
\end{tabular}
\label{tab-ghcar-oc}
\end{table}

\paragraph*{Binarity of GH~Car}
\label{ghcar-radvel}

\begin{figure}
\includegraphics[height=58mm, angle=0]{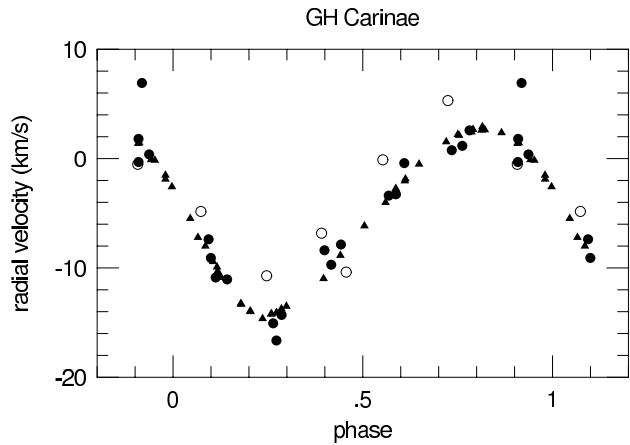}
\caption{Radial velocity phase curve of GH~Carinae. 
Filled circles represent data from 2004-2005, open
circles denote data from 2006, while the \textit{CORALIE}
data obtained in 2011-2012 are marked as triangles}
\label{fig-ghcar-vrad}
\end{figure}

There are no published radial velocity data for this 
bright Cepheid. The phase diagram of our RV observations 
is plotted in Fig.~\ref{fig-ghcar-vrad}.
The observational data have been folded on the period given
by the ephemeris in Eq.~\ref{ghcar-ephemeris}. The zero phase
has been arbitrarily chosen at JD\,2\,400\,000 (similarly to
all phase curves in this paper). Figure~\ref{fig-ghcar-vrad}
clearly shows a vertical shift between the mean values valid for 
2004-2005 and 2006. For the first season, the $\gamma$-velocity
(the mean RV averaged over a pulsational cycle)
was $-$4.6 km\,s$^{-1}$, while one year later it became 
$-$3.5 km\,s$^{-1}$, while the \textit{CORALIE} data result 
in the value of $-$5.3 km\,s$^{-1}$ for the $\gamma$-velocity.
Though the difference is small, homogeneity of the data and 
the identical treatment is a guarantee that the shift is not 
an artifact of the analysis. The individual data are listed 
in Tables~\ref{tab-ghcar-data}--\ref{tab-ghcar-coralie-data}.

Spectroscopic binarity of GH~Car has to be verified by further
observations.

\begin{table}
\caption{New RV values of GH Carinae from the SSO spectra.
This is only a portion of the full version 
available online only}
\begin{tabular}{lr}
\hline
\noalign{\vskip 0.2mm}
JD$_{\odot}$ & $v_{\rm rad}$ \  \\
2\,400\,000 + &(km\,s$^{-1}$)\\
\noalign{\vskip 0.2mm}
\hline
\noalign{\vskip 0.2mm}
53364.2431&$-$8.4\\
53367.2163& 6.9\\
53368.2542&$-$9.1\\
53369.2434&$-$16.6\\
53451.0951&$-$3.4\\
\noalign{\vskip 0.2mm}
\hline
\end{tabular}
\label{tab-ghcar-data}
\end{table}

\begin{table}
\caption{New \textit{CORALIE} velocities of GH Carinae.
This is only a portion of the full version 
available online only}
\begin{tabular}{lr}
\hline
\noalign{\vskip 0.2mm}
JD$_{\odot}$ & $v_{\rm rad}$ \  \\
2\,400\,000 + &(km\,s$^{-1}$)\\
\noalign{\vskip 0.2mm}
\hline
\noalign{\vskip 0.2mm}
55652.8343&	$-$9.9\\
55653.8063&	$-$13.7\\
55654.6963&	$-$8.9\\
55655.6685&	$-$2.0\\
55656.6753&	2.4\\
\noalign{\vskip 0.2mm}
\hline
\end{tabular}
\label{tab-ghcar-coralie-data}
\end{table}

\subsection{V419~Centauri}
\label{v419cen}

\paragraph*{Accurate value of the pulsation period}
\label{v419cen-period}

The brightness variability of V419~Cen (HD\,100148, $\langle V \rangle
= 8.19$\,mag.) was revealed by \citet{OO37}.
This Cepheid also pulsates in the first overtone mode, so the 
$O-C$ diagram was also constructed for the moments of the median
brightness on the ascending branch, similarly to the case of
GH~Car (see Sect.~\ref{ghcar-period}).

\begin{table}
\caption{$O-C$ values of V419~Centauri (see the description in 
Sect.~\ref{ghcar-period})}
\begin{tabular}{l@{\hskip2mm}r@{\hskip2mm}r@{\hskip2mm}c@{\hskip2mm}l}
\hline
\noalign{\vskip 0.2mm}
JD$_{\odot}$ & $E\ $ & $O-C$ & $W$ & Data source\\
2\,400\,000 + &&&\\
\noalign{\vskip 0.2mm}
\hline
\noalign{\vskip 0.2mm}
34817.2512 &$-$3185 &+0.0431 &2& \citet{Wetal58}\\
35219.2722 &$-$3112 &+0.0441 &3& \citet{I61}\\
40285.8001 &$-$2192 &+0.0188 &3& \citet{S70}\\
40737.4108 &$-$2110 &+0.0454 &3& \citet{P76}\\
42896.1636 &$-$1718 &+0.0060 &3& \citet{D77}\\
44300.3547 &$-$1463&$-$0.1193 &2& \citet{B08}\\
44630.7521 &$-$1403&$-$0.1492 &2& \citet{E85}\\
48166.4174 &$-$761&$-$0.0569 &3& {\it Hipparcos} \citep{ESA97}\\
48689.5720 &$-$666&$-$0.0790 &3& {\it Hipparcos} \citep{ESA97}\\
49543.1967 &$-$511&$-$0.0583 &2& \citet{B08}\\
49813.0946 &$-$462&$-$0.0095 &3& \citet{B08}\\
50385.8636 &$-$358 &+0.0187 &2& \citet{B08}\\
50573.1155 &$-$324 &+0.0285 &3& \citet{B08}\\
50903.5445 &$-$264 &+0.0301 &3& \citet{B08}\\
51261.4846 &$-$199 &+0.0072 &3& \citet{B08}\\
51647.0071 &$-$129 &+0.0311 &3& \citet{B08}\\
51955.3943  &$-$73 &+0.0194 &3& ASAS \citep{P02}\\
51960.9164  &$-$72 &+0.0344 &3& \citet{B08}\\
52126.0927  &$-$42 &+0.0030 &3& ASAS \citep{P02}\\
52357.3861 & 0&$-$0.0088 &3& \citet{B08}\\ 
52467.5557 & 20 & +0.0183 &3& ASAS \citep{P02}\\
52643.7458 & 52&$-$0.0195 &3& \citet{B08}\\
52731.8585 & 68&$-$0.0208 &3& ASAS \citep{P02}\\
53012.7265 & 119&$-$0.0160 &3& \citet{B08}\\
53062.2908 & 128&$-$0.0158 &3& ASAS \citep{P02}\\
53475.3737 & 203 &+0.0328 &3& ASAS \citep{P02}\\
53524.8954 & 212&$-$0.0096 &2& {\it INTEGRAL} OMC\\
53789.2626 & 260 &+0.0157 &3& ASAS \citep{P02}\\
54136.1005 & 323&$-$0.0951 &1& {\it INTEGRAL} OMC\\
54191.2873 & 333 &+0.0204 &3& ASAS \citep{P02}\\
54587.8324 & 405 &+0.0527 &3& ASAS \citep{P02}\\
54918.2426 & 465 &+0.0355 &3& ASAS \citep{P02}\\
\noalign{\vskip 0.2mm}
\hline
\end{tabular}
\label{tab-v419cen-oc}
\end{table}

\begin{figure}
\includegraphics[height=44mm, angle=0]{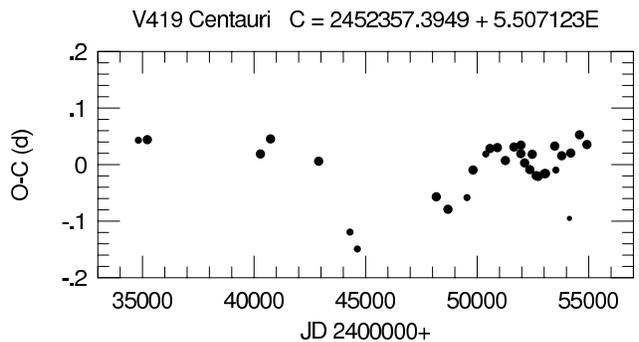}
\caption{$O-C$ diagram of V419~Cen. The plot can be
approximated by a constant period but a parabolic pattern 
indicating a continuously increasing period cannot be excluded}
\label{fig-v419cen-oc}
\end{figure}

The $O-C$ diagram of V419~Cen based on the $O-C$ values listed
in Table~\ref{tab-v419cen-oc} is shown plotted in 
Fig.~\ref{fig-v419cen-oc}. The plot can be approximated by a constant 
period by the ephemeris for the moments of median brightness 
on the ascending branch:
\vspace{-1mm}
\begin{equation}
C = 2\,452\,357.3949 + 5.507\,123{\times}E 
\label{v419cen-ephemeris}
\end{equation}
\vspace{-3mm}
$\phantom{mmmmm}\pm0.0053\phantom{}\pm 0.000\,005$

\noindent However, a parabolic pattern indicating a continuously 
increasing period cannot be excluded. For the proper phasing of
the RV data it is important that the $O-C$ differences
are about zero for each epoch when radial velocity data were 
obtained.

\paragraph*{Binarity of V419~Cen}
\label{v419cen-bin}

\begin{table}
\caption{New RV values of V419 Centauri from the SSO spectra.
This is only a portion of the full version 
available online only}
\begin{tabular}{lr}
\hline
\noalign{\vskip 0.2mm}
JD$_{\odot}$ & $v_{\rm rad}$  \ \\
2\,400\,000 + &(km\,s$^{-1}$)\\
\noalign{\vskip 0.2mm}
\hline
\noalign{\vskip 0.2mm}
53369.2636&$-$1.8\\
53451.1179&$-$4.8\\
53452.0995& 3.9\\
53453.1895&$-$1.5\\
53454.1215&$-$7.9\\
\noalign{\vskip 0.2mm}
\hline
\end{tabular}
\label{tab-v419cen-data}
\end{table}

\begin{table}
\caption{$\gamma$-velocities of V419~Centauri}
\begin{tabular}{lccl}
\hline
\noalign{\vskip 0.2mm}
Mid-JD & $v_{\gamma}$ & $\sigma$ & Data source \\
2\,400\,000+ & (km\,s$^{-1}$)& (km\,s$^{-1}$) & \\
\noalign{\vskip 0.2mm}
\hline
\noalign{\vskip 0.2mm}
34129 & $-$16.40& 1.3 & \cite{S55}\\
40550 & $-$11.70& 1.5 & \cite{LE80}\\
53474 & $-$5.53 & 0.3 & present paper\\
53800 & $-$0.96 & 0.5 & present paper\\
\noalign{\vskip 0.2mm}
\hline
\end{tabular}
\label{tab-v419cen-vgamma}
\end{table}

\begin{figure}
\includegraphics[height=58mm, angle=0]{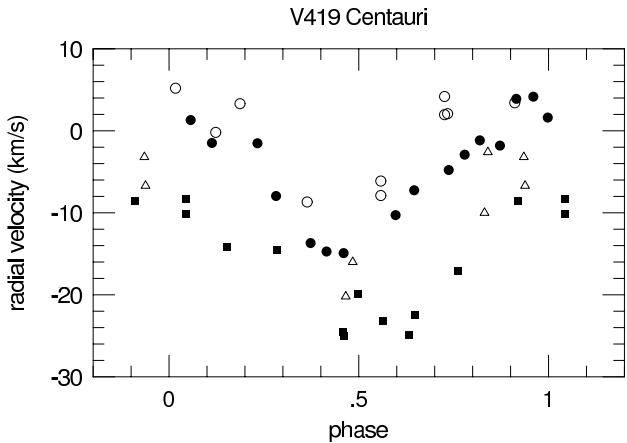}
\caption{Merged RV phase curve of V419~Cen.
There is a striking difference between the 
$\gamma$-velocities valid for the epoch of Stibbs' 
(\citeyear{S55}) and Lloyd Evans' (\citeyear{LE80}) data 
(denoted as filled squares and empty triangles, respectively) 
and our recent data (denoted by circles, see the text)}
\label{fig-v419cen-vrad}
\end{figure}

\begin{figure}
\includegraphics[height=45mm, angle=0]{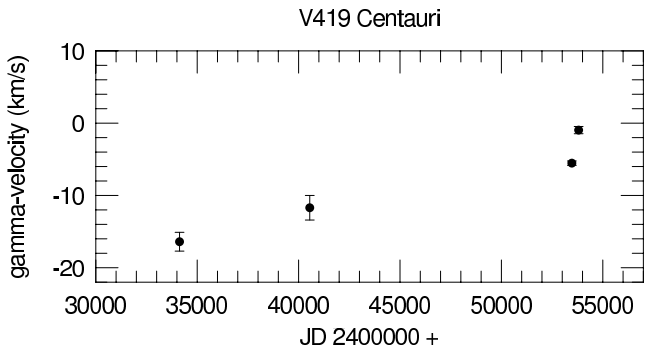}
\caption{Temporal shift in the $\gamma$-velocity of V419~Cen}
\label{fig-v419cen-vgamma}
\end{figure}

All RV data (including the new ones listed 
in Table~\ref{tab-v419cen-data}) have been folded on the 
accurate pulsation period taken from the ephemeris given in 
Eq.~\ref{v419cen-ephemeris}, so the different data series have 
been correctly phased with respect to each other. The merged 
RV phase curve is plotted in Fig.~\ref{fig-v419cen-vrad}.
The individual data series are denoted with different symbols:
filled squares - radial velocities from 1952 by \citet{S55};
empty triangles - data by \citet{LE80} from 1969-1970; 
filled and empty circles - our 2004-2005 and 2006 data, respectively.

Variability in the $\gamma$-velocity is striking. Systematic 
errors can be excluded. Although our 2006 data are shifted to 
a larger value of the $\gamma$-velocity, similarly to the case
of GH~Car (see Sect.~\ref{ghcar-radvel}), other new spectroscopic
binaries and dozens of Cepheids in our sample with non-varying
$\gamma$-velocities indicate stability of the equipment and 
reliability of the data reduction. Another piece of evidence in 
favour of the intrinsic variability of the $\gamma$-velocity is that 
both \citet{S55} and \citet{LE80} used the same spectrograph
during their observations.

To have a clearer picture, the $\gamma$-velocities (together with 
their uncertainties) are listed in Table~\ref{tab-v419cen-vgamma} 
and also plotted in Fig.~\ref{fig-v419cen-vgamma}. The last two
points (which are the most accurate ones), i.e., the shift 
between 2004-2005 and 2006 data implies an orbital period of 
several years instead of several decades suggested by the whole 
pattern of the plot.

\subsection{V898~Centauri}
\label{v898cen}

\paragraph*{Accurate value of the pulsation period}
\label{v898cen-period}

The brightness variability of V898~Cen (HD\,97317, $\langle V \rangle
= 8.00$\,mag.) was revealed by \citet{Setal64}. This is also a
low amplitude Cepheid pulsating in the first overtone mode. 
The first reliable photometric data were only obtained during the 
{\it Hipparcos} space astrometry mission \citep{ESA97}. Later on 
Berdnikov and his co-workers followed the photometric behaviour 
of V898~Cen \citep{B08}.

The $O-C$ diagram of V898~Cen was constructed for the
moments of median brightness on the ascending branch
(see Table~\ref{tab-v898cen-oc}). The weighted least
squares fit to the $O-C$ values resulted in the
ephemeris:
\vspace{-1mm}
\begin{equation}
C = 2\,452\,353.8356 + 3.527\,310{\times}E 
\label{v898cen-ephemeris}
\end{equation}
\vspace{-3mm}
$\phantom{mmmmm}\pm0.0052\phantom{}\pm 0.000\,008$

\begin{figure}
\includegraphics[height=44mm, angle=0]{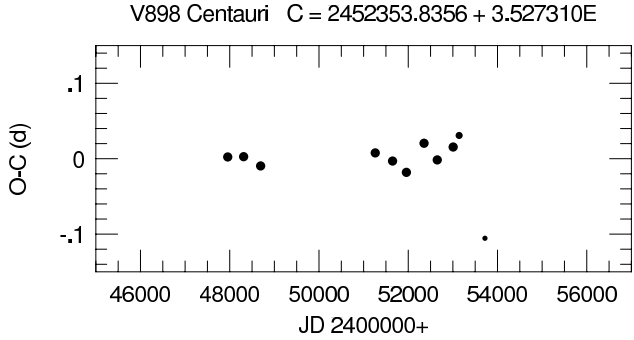}
\caption{$O-C$ diagram of V898~Cen. The plot can be
approximated by a constant period}
\label{fig-v898cen-oc}
\end{figure}

\noindent The $O-C$ diagram of V898~Cen plotted in
Fig.~\ref{fig-v898cen-oc} indicates constancy of the
pulsation period.

\paragraph*{Binarity of V898~Cen}
\label{v898cen-bin}

The Cepheid variable V898~Cen has been neglected from the
point of view of spectroscopy, as well. A spectral type of
F3III has been assigned to it in the SIMBAD data base which
is atypical of a Cepheid (and probably erroneous). Cepheids
are supergiants of Iab or Ib luminosity class and their
short period representatives have a late F spectral type.
Ironically there is a single RV data, $-2.4\pm 2.4$
km\,s$^{-1}$ (which is an average of two measurements)
published by \citet{Netal04} in their catalog of 14000 {\em dwarf}
stars of F and G spectral types. However, the epoch of these 
particular observations has remained unknown. Therefore, 
our data provide a first epoch RV phase curve.

\begin{table}
\caption{$O-C$ values of V898~Centauri (description of the columns
is given in Sect.~\ref{ghcar-period})}
\begin{tabular}{l@{\hskip2mm}r@{\hskip2mm}r@{\hskip2mm}c@{\hskip2mm}l}
\hline
\noalign{\vskip 0.2mm}
JD$_{\odot}$ & $E\ $ & $O-C$ & $W$ & Data source\\
2\,400\,000 + &&&\\
\noalign{\vskip 0.2mm}
\hline
\noalign{\vskip 0.2mm}
47955.2823 &$-$1247 &+0.0023 &3& {\it Hipparcos} \citep{ESA97}\\
48311.5410 &$-$1146 &+0.0027 &3& {\it Hipparcos} \citep{ESA97}\\
48692.4782 &$-$1038&$-$0.0096 &3& {\it Hipparcos} \citep{ESA97}\\
51260.3772 &$-$310 &+0.0077 &3& \citet{B08}\\
51648.3705 &$-$200&$-$0.0031 &3& \citet{B08}\\
51958.7588 &$-$112&$-$0.0181 &3& \citet{B08}\\
52353.8561 &0 &+0.0205 &3& \citet{B08}\\
52650.1280 &84&$-$0.0016 &3& \citet{B08}\\
53006.4034 &185 &+0.0155 &3& \citet{B08}\\
53140.4565 &223 &+0.0308 &2&  {\it INTEGRAL} OMC\\
53718.7991 &387&$-$0.1055 &1&  {\it INTEGRAL} OMC\\
\noalign{\vskip 0.2mm}
\hline
\end{tabular}
\label{tab-v898cen-oc}
\end{table}

\begin{figure}
\includegraphics[height=42mm, angle=0]{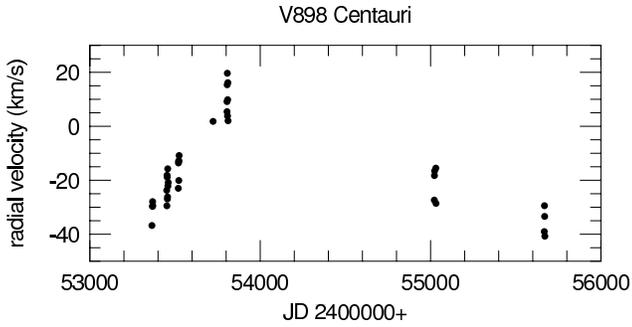}
\caption{Merged RV curve of V898~Cen
}
\label{fig-v898cen-vradJD}
\end{figure}

\begin{figure}
\includegraphics[height=55mm, angle=0]{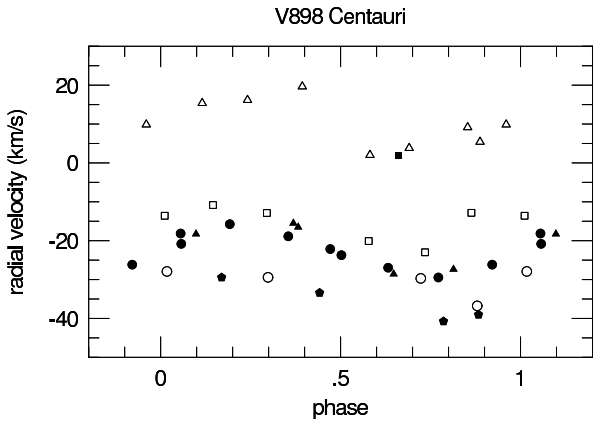}
\caption{Radial velocity phase curve of V898~Cen. 
Various symbols refer to different observational runs 
(see the text)}
\label{fig-v898cen-vrad}
\end{figure}

\begin{figure}
\includegraphics[height=42mm, angle=0]{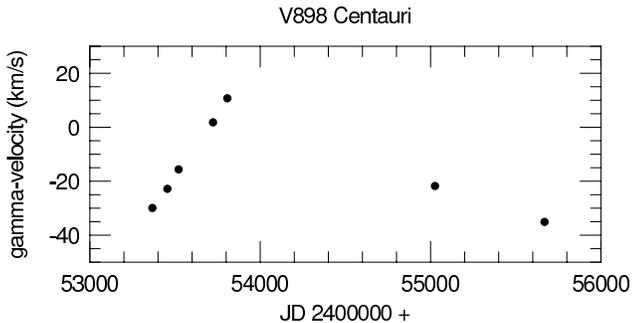}
\caption{Temporal drift in the $\gamma$-velocity
of V898~Centauri}
\label{fig-v898cen-vgamma}
\end{figure}

\begin{table}
\caption{New RV values of V898 Centauri from the SSO spectra.
This is only a portion of the full version 
available online only}
\begin{tabular}{lr}
\hline
\noalign{\vskip 0.2mm}
JD$_{\odot}$ & $v_{\rm rad}$ \  \\
2\,400\,000 + &(km\,s$^{-1}$)\\
\noalign{\vskip 0.2mm}
\hline
\noalign{\vskip 0.2mm}
53364.2485&$-$36.8\\
53367.2225&$-$29.7\\
53368.2618&$-$27.9\\
53369.2529&$-$29.4\\
53451.1003&$-$23.7\\
\noalign{\vskip 0.2mm}
\hline
\end{tabular}
\label{tab-v898cen-data}
\end{table}

\begin{table}
\caption{New \textit{FEROS} velocities of V898 Centauri}
\begin{tabular}{lr}
\hline
\noalign{\vskip 0.2mm}
JD$_{\odot}$ & $v_{\rm rad}$ \  \\
2\,400\,000 + &(km\,s$^{-1}$)\\
\noalign{\vskip 0.2mm}
\hline
\noalign{\vskip 0.2mm}
55667.5948& $-$39.02\\
55668.6044& $-$29.44\\
55669.5644& $-$33.40\\
55670.7789& $-$40.74\\
\noalign{\vskip 0.2mm}
\hline
\end{tabular}
\label{tab-v898cen-feros-data}
\end{table}

\begin{table}
\caption{$\gamma$-velocities of V898~Centauri}
\begin{tabular}{lccl}
\hline
\noalign{\vskip 0.2mm}
Mid-JD & $v_{\gamma}$ & $\sigma$ & Data source \\
2\,400\,000+ & (km\,s$^{-1}$)& (km\,s$^{-1}$) & \\
\noalign{\vskip 0.2mm}
\hline
\noalign{\vskip 0.2mm}
53367.7 & $-$29.9& 0.5 & present paper\\
53455.5 & $-$22.8& 0.4 & present paper\\
53521.4 & $-$15.6& 0.5 & present paper\\
53723.3 & \ \ 1.8 & 0.6 & present paper\\
53807.5 & \ 10.7& 0.4 & present paper\\
55026.3 & \ $-$21.8& 0.5 & present paper\\
55669.2 & \ $-$35.1& 0.5 & present paper\\
\noalign{\vskip 0.2mm}
\hline
\end{tabular}
\label{tab-v898cen-vgamma}
\end{table}

It became obvious already in the first observing season that
an orbital effect is superimposed on the RV
variations of pulsational origin (see the left part of
Fig.~~\ref{fig-v898cen-vradJD}). Therefore several spectra of 
V898~Cen have been taken in 2009 and 2011, as well. Our
individual RV data are listed in Tables~\ref{tab-v898cen-data}
and \ref{tab-v898cen-feros-data}. Based on these data, the RV 
phase curve has been constructed using the 3.527310~d 
pulsation period appearing in Eq.~\ref{v898cen-ephemeris}. 
The wide scatter in this phase curve plotted in 
Fig.~\ref{fig-v898cen-vrad} corresponds to a variable 
$\gamma$-velocity. The data series has been split into seven 
segments, denoted by different symbols in Fig.~\ref{fig-v898cen-vrad}: 
empty circles - Dec. 2004; filled circles - March 2005; 
empty squares - May-June 2005; filled square - Dec. 2005; 
empty triangles - March 2006; filled triangles - July 2009;
filled pentagons - April 2011.

The $\gamma$-velocities determined from each data segment
are listed in Table~\ref{tab-v898cen-vgamma} and are plotted
in Fig.~\ref{fig-v898cen-vgamma}. This latter plot implies
that V898~Cen is a new spectroscopic binary and the orbital 
period is about 2000 or 3000 days, depending on whether the
most negative value of the $\gamma$-velocity occurred before
or after our measurements in 2011. The pattern of the points
implies a non-sinusoidal shape of the orbital velocity
phase curve, with much steeper ascending branch than descending
one. This is a strong indication of an eccentric orbit, however
far more seasons of observations  are needed before attempting 
to derive accurate orbital elements from the $\gamma$-velocity 
variations. The most important feature of Fig.~\ref{fig-v898cen-vgamma}
is the large amplitude of the orbital velocity  variations: 
it exceeds 40 km\,s$^{-1}$. The `recorder' among the known 
binary systems involving a Cepheid component has been the 
system of SU~Cygni with a peak-to-peak orbital amplitude of 
60 km\,s$^{-1}$ \citep[see the on-line data base of
Cepheids in binary systems described by][]{Sz03a}.

The orbital motion of the Cepheid component around the center
of mass in the binary system may cause a light-time effect in
the $O-C$ diagram of the given variable star. In the case of
V898~Cen the wave-like pattern characteristic of the light-time
effect cannot be detected yet (see Fig.~\ref{fig-v898cen-oc}).

To provide reliable values for the physical properties of 
this bright Cepheid, our \textit{FEROS} spectra were analysed 
in detail. The parameters $T_{\rm eff}$, $\log g$, [M/H], and 
$v \sin i$ were determined by searching for the best
fitting model in the synthetic spectrum library of \citet{Metal05}
using a standard $\chi^2$ procedure. To derive the parameters and 
their errors we applied the following method: The model spectra 
were ordered according to their calculated $\chi^2$ value, then 
we selected all with number less than $1.05\,\chi_{\rm min}^2$ 
and adopted the means and the standard deviations of this sample 
as values and errors of the parameters.

The best fitting values are as follows:\\
\noindent$T_{\rm eff} = 5950 \pm 380$ K,\\
\noindent$\log g = 1.2 \pm ${\boldmath$0.7$},\\
\noindent[M/H]$ = -0.4 \pm 0.2$\\
\noindent$v \sin i = 2 \pm 2$.

The effective temperature obtained by us corresponds to an F9
spectral type supergiant star, a typical value of a short-period
Cepheid. Note, however, that these values are preliminary ones.
The large number of SSO spectra of all 40 target Cepheids will be 
analysed for obtaining physical properties with smaller uncertainties.

\subsection{AD~Puppis}
\label{adpup}

\paragraph*{Accurate value of the pulsation period}
\label{adpup-period}

The brightness variability of AD~Pup (HD\,63446, $\langle V \rangle
= 9.91$\,mag.) was revealed by Hertzsprung \citep{W35}. This is the
longest period Cepheid in this paper and it has been frequently 
observed from the 1950s. Long-period Cepheids are usually fundamental 
pulsators and they oscillate with a large amplitude. In their case,
the $O-C$ analysis is based on the moments of brightness maxima.

\begin{figure}
\includegraphics[height=44mm, angle=0]{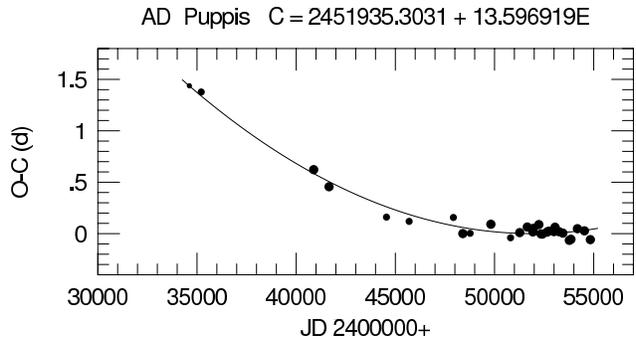}
\caption{$O-C$ diagram of AD~Pup. The plot can be
approximated by a parabola indicating a continuously
increasing pulsation period}
\label{fig-adpup-oc}
\end{figure}

The $O-C$ differences of AD~Puppis are listed in Table~\ref{tab-adpup-oc}.
These values have been obtained by the following ephemeris:
\vspace{-1mm}
\begin{equation}
C = 2\,451\,935.3031 + 13.596\,919{\times}E
\label{adpup-ephemeris}
\end{equation}
\vspace{-3mm}
$\phantom{mmmmm}\pm0.0065\phantom{\,}\pm 0.000\,040$

\noindent which contains the constant and linear terms of the
weighted parabolic fit to the $O-C$ values. The parabolic nature
of the $O-C$ diagram is clearly seen in Fig.~\ref{fig-adpup-oc}.

\begin{table}
\caption{$O-C$ values of AD~Puppis (description of the columns
is given in Sect.~\ref{ghcar-period} but the first column contains
the moments of brightness maxima)}
\begin{tabular}{l@{\hskip2mm}r@{\hskip2mm}r@{\hskip2mm}c@{\hskip2mm}l}
\hline
\noalign{\vskip 0.2mm}
JD$_{\odot}$ & $E\ $ & $O-C$ & $W$ & Data source\\
2\,400\,000 + &&&\\
\noalign{\vskip 0.2mm}
\hline
\noalign{\vskip 0.2mm}
34614.2663 &$-$1274 & +1.4380 & 1 & \citet{Wetal58}\\
35212.4702 &$-$1230 & +1.3775 & 2 & \citet{I61}\\
40881.6304 &$-$813 & +0.6224 & 3 & \citet{P76}\\
41656.4884 &$-$756 & +0.4561 & 3 & \citet{M75}\\
44552.3367 &$-$543 & +0.1606 & 2 & \citet{E83}\\
45694.4366 &$-$459 & +0.1193 & 2 & \citet{B08}\\
47924.3692 &$-$295 & +0.1572 & 2 & {\it Hipparcos} \citep{ESA97}\\
48400.1048 &$-$260 & +0.0006 & 3 & {\it Hipparcos} \citep{ESA97}\\
48767.2244 &$-$233 & +0.0034 & 2 & {\it Hipparcos} \citep{ESA97}\\
49814.2750 &$-$156 & +0.0913 & 3 & \citet{B08} \\
50806.7186 & $-$83 &$-$0.0402 & 2 & \citet{B08}\\
51269.0641 & $-$49 & +0.0100 & 3 & \citet{B08}\\
51649.8321 & $-$21 & +0.0643 & 3 & \citet{B08}\\
51935.3197 &    0 & +0.0166 & 3 & ASAS \citep{P02}\\
51962.5484 &    2 & +0.0515 & 3 & \citet{B08}\\
52234.5236 &   22 & +0.0883 & 3 & ASAS \citep{P02}\\
52343.2076 &   30 &$-$0.0031 & 3 & \citet{B08}\\
52411.1912 &   35 &$-$0.0041 & 3 & ASAS \citep{P02}\\
52642.3572 &   52 & +0.0143 & 3 & \citet{B08}\\
52723.9484 &   58 & +0.0240 & 3 & ASAS \citep{P02}\\
52995.8827 &   78 & +0.0199 & 3 & \citet{B08} \\
53050.3131 &   82 & +0.0626 & 3 & ASAS \citep{P02}\\
53254.2231 &   97 & +0.0189 & 3 & {\it INTEGRAL} OMC\\
53444.5663 &  111 & +0.0052 & 3 & ASAS \citep{P02}\\
53770.8223 &  135 &$-$0.0649 & 3 & ASAS \citep{P02}\\
53852.4136 &  141 &$-$0.0551 & 3 & ASAS \citep{P02}\\
54178.8427 &  165 & +0.0480 & 3 & ASAS \citep{P02}\\
54532.3418 &  191 & +0.0272 & 3 & ASAS \citep{P02}\\
54831.3885 &  213 &$-$0.0583 & 3 & ASAS \citep{P02}\\
\noalign{\vskip 0.2mm}
\hline
\end{tabular}
\label{tab-adpup-oc}
\end{table}

This parabolic trend corresponds to a continuous period increase
of $(1.7 \pm 0.09)\times 10^{-6}$ d/cycle, i.e., 
$\Delta P = 0.004567$ d/century.
This tiny but non-negligible period increase has been caused
by stellar evolution: while the Cepheid crosses the instability
region towards lower temperatures in the Hertzsprung-Russell 
diagram, its pulsation period is increasing. 
Continuous period variations (of either sign) often occur in 
the pulsation of long-period Cepheids \citep{Sz83}.

The pattern of fluctuations around the fitted parabola
shows a wavy nature with a characteristic period of about 50 years 
as if it were a light-time effect.

\paragraph*{Binarity of AD~Pup}
\label{adpup-bin}

\begin{figure}
\includegraphics[height=60mm, angle=0]{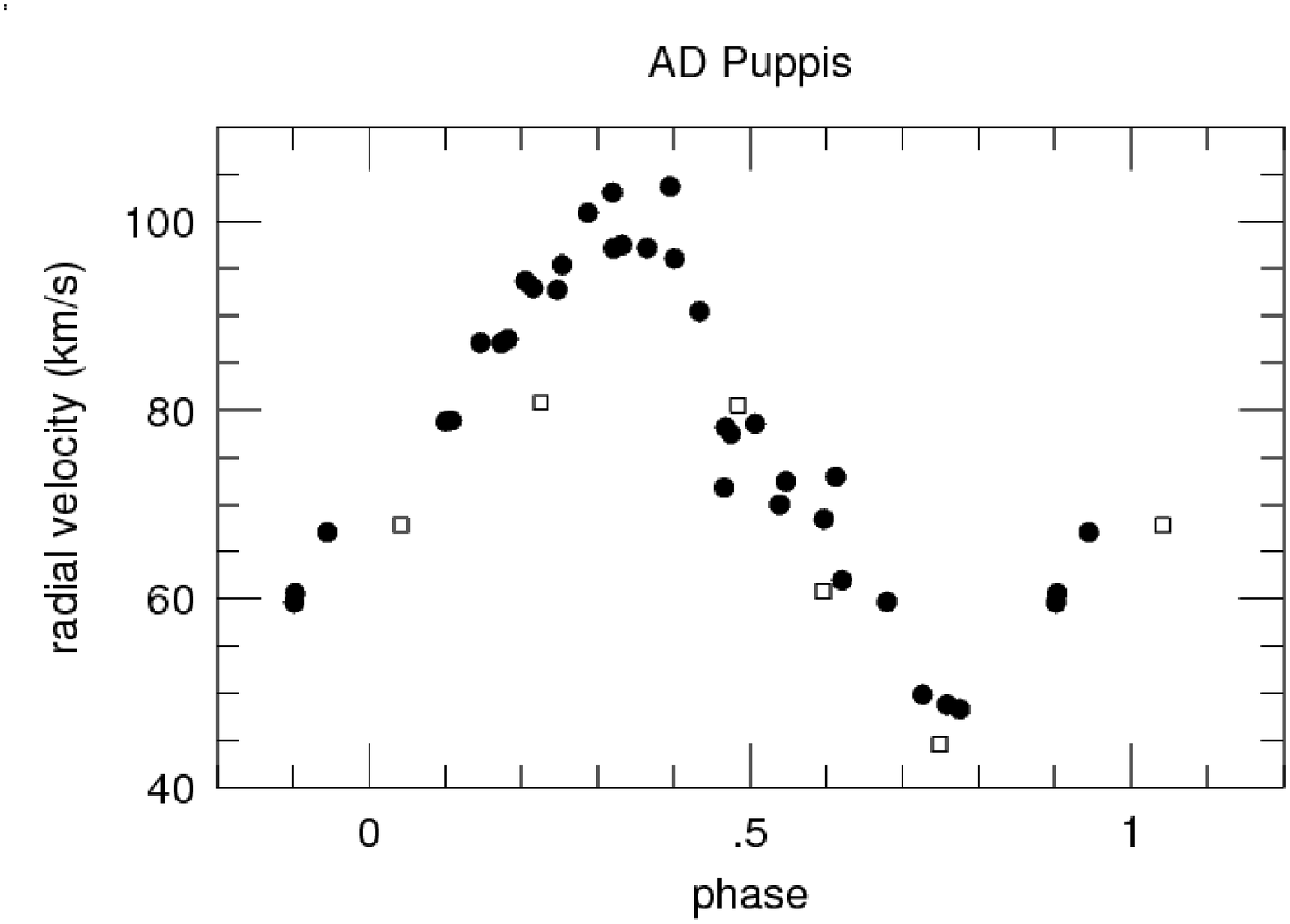}
\caption{Merged RV phase curve of AD~Pup.
There is an obvious shift between the 
$\gamma$-velocities valid for the epoch of Joy's 
(\citeyear{J37}) data (denoted by empty squares) and our 
data (filled circles)}
\label{fig-adpup-vrad}
\end{figure}

\begin{table}
\caption{New RV values of AD Puppis from the SSO spectra.
This is only a portion of the full version available online only}
\begin{tabular}{lr}
\hline
\noalign{\vskip 0.2mm}
JD$_{\odot}$ & $v_{\rm rad}$  \  \\
2\,400\,000 + &(km\,s$^{-1}$)\\
\noalign{\vskip 0.2mm}
\hline
\noalign{\vskip 0.2mm}
53304.2658& 103.1\\
53306.2580& 71.8\\
53307.2487& 70.0\\
53308.2504& 72.9\\
53309.1650& 59.7\\
\noalign{\vskip 0.2mm}
\hline
\end{tabular}
\label{tab-adpup-data}
\end{table}

The earlier RV data by \citet{J37} imply a
significantly different $\gamma$-velocity (66.5 km\,s$^{-1}$)
than our recent ones (74.0 km\,s$^{-1}$) in spite of the uncertainty 
of his individual data as large as 4 km\,s$^{-1}$. 
Because the zero point of Joy's system is reliable, as discussed 
by \citet{Sz96}, there is no systematic difference of instrumental 
or data treatment origin between Joy's and the more recent 
observational series. The only plausible explanation for the shift 
in the $\gamma$-velocity is the orbital motion in a binary system 
superimposed on the pulsational RV changes. The shift in the 
$\gamma$-velocity is obvious in the phase diagram of the radial 
velocities of AD~Puppis plotted in Fig.~\ref{fig-adpup-vrad} 
where Joy's data are denoted with empty squares, while our data 
are represented with filled circles. The $\gamma$-velocity of AD~Pup 
did not change noticeably during the interval of our observations. 
Our RV data (listed in Table~\ref{tab-adpup-data} have been folded 
with the period as given in the ephemeris in Eq.~\ref{adpup-ephemeris}. 
Joy's data have been phased with the same period but a proper 
correction has been applied to correct for the phase shift due to 
the parabolic $O-C$ graph.

The smaller value of the $\gamma$-velocity determined from Joy's
(\citeyear{J37}) data is in a qualitative agreement with the 
with the wave-like pattern superimposed on the fitted parabola
in Fig.~\ref{fig-adpup-oc}. In this particular case the light-time
effect interpretation implies a 50-year-long orbital period.

\subsection{AY~Sagittarii}
\label{aysgr}

\paragraph*{Accurate value of the pulsation period}
\label{aysgr-period}

\begin{figure}
\includegraphics[height=44mm, angle=0]{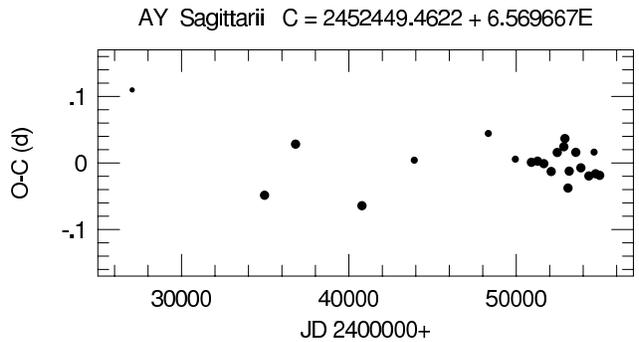}
\caption{$O-C$ diagram of AY~Sgr. The plot can be
approximated by a constant period}
\label{fig-aysgr-oc}
\end{figure}

The brightness variability of AY~Sgr (HIP\,90110, $\langle V \rangle
= 10.55$\,mag.) was revealed by Henrietta Leavitt \citep{P04}. 
\citet{H23} determined the pulsation period to be 6.74426 days 
from his unpublished visual observations made in 1917-1918. 
Interestingly enough, this period is about 3\% longer than the
value deduced from the $O-C$ diagram (see Table~\ref{tab-aysgr-oc} 
and Fig.~\ref{fig-aysgr-oc}). Such a strong period change, if it
really happened, is unprecedented among classical Cepheids.

\begin{table}
\caption{$O-C$ values of AY~Sagittarii (description of the columns
is given in Sect.~\ref{ghcar-period} but the first column contains
the moments of brightness maxima)}
\begin{tabular}{l@{\hskip2mm}r@{\hskip2mm}r@{\hskip2mm}c@{\hskip2mm}l}
\hline
\noalign{\vskip 0.2mm}
JD$_{\odot}$ & $E\ $ & $O-C$ & $W$ & Data source\\
2\,400\,000 + &&&\\
\noalign{\vskip 0.2mm}
\hline
\noalign{\vskip 0.2mm}
21425.45  &$-$4722 &$-$2.04& - &\citet{H23}\\
27051.2394&$-$3866 & 0.1098 &1& \citet{FK53}\\
34960.9601&$-$2662 &$-$0.0485 &3& \citet{Wetal58}\\
36813.6830&$-$2380 & +0.0283 &3& \citet{Wetal60}\\
40781.6694&$-$1776 &$-$0.0642 &3& \citet{P76}\\
43908.8993&$-$1300 & +0.0042 &2& \citet{H80}\\
48330.3254 &$-$627 & +0.0444 &2& {\it Hipparcos} \citep{ESA97}\\
49946.4248 &$-$381 & +0.0057 &2& \citet{B08}\\
50905.5915 &$-$235 & +0.0010 &3& \citet{B08}\\
51273.5280 &$-$179 & +0.0027 &3& \citet{B08}\\
51647.9618 &$-$122&$-$0.0010 &3& \citet{B08}\\
52088.1177 &$-$55&$-$0.0128 &3& ASAS \citep{P02}\\
52449.4780 &   0 & +0.0158 &3& ASAS \citep{P02} \\
52843.6666 &  60 & +0.0244 &3& ASAS \citep{P02}\\
52909.3756 &  70 & +0.0367 &3& {\it INTEGRAL} OMC\\
53093.2519 &  98 &$-$0.0377 &3& \citet{B08}\\
53165.5436 &  109&$-$0.0123 &3& ASAS \citep{P02}\\
53559.7519 &  169 & +0.0160 &3& ASAS \citep{P02}\\
53861.9333 &  215&$-$0.0073 &3& ASAS \citep{P02}\\
54341.5068 &  288&$-$0.0195 &3& ASAS \citep{P02}\\
54650.3169 &  335 & +0.0163 &2& {\it INTEGRAL} OMC\\
54742.2598 &  349&$-$0.0162 &3& ASAS \citep{P02}\\
54985.3351 &  386&$-$0.0186 &3& ASAS \citep{P02}\\
\noalign{\vskip 0.2mm}
\hline
\end{tabular}
\label{tab-aysgr-oc}
\end{table}

AY~Sgr is a fundamental mode pulsator, so its $O-C$ diagram 
has been constructed based on the moments of brightness maxima.
The tabulated $O-C$ values have been calculated by using
the ephemeris:
\vspace{-1mm}
\begin{equation}
C = 2\,452\,449.4622 + 6.569\,667{\times}E 
\label{aysgr-ephemeris}
\end{equation}
\vspace{-3mm}
$\phantom{mmmmm}\pm0.0044\phantom{}\pm 0.000\,004$

\noindent based on applying a weighted linear least
squares fit to the $O-C$ differences.

As is seen in Fig.~\ref{fig-aysgr-oc}, the pulsation period
of AY~Sgr has remained practically constant over decades.

\paragraph*{Binarity of AY~Sgr}
\label{aysgr-bin}

\begin{figure}
\includegraphics[height=58mm, angle=0]{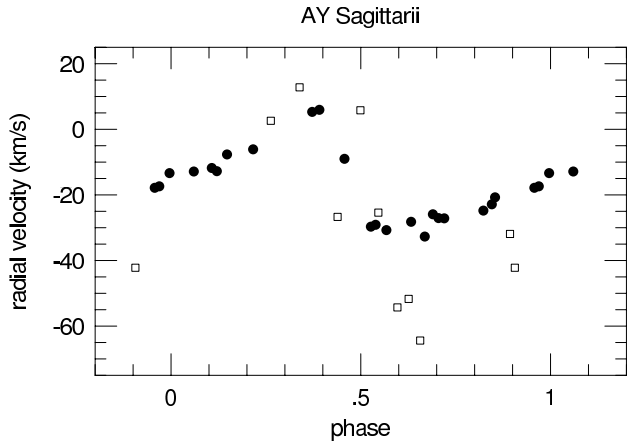}
\caption{Merged RV phase curve of AY~Sgr.
There is a striking difference between the 
$\gamma$-velocities valid for the epoch of Joy's 
(\citeyear{J37}) data (denoted as empty squares) and our 
recent data (filled circles)}
\label{fig-aysgr-vrad}
\end{figure}

\begin{table}
\caption{New RV values of AY Sagittarii from the SSO spectra.
This is only a portion of the full version 
available online only}
\begin{tabular}{lr}
\hline
\noalign{\vskip 0.2mm}
JD$_{\odot}$ & $v_{\rm rad}$ \  \\
2\,400\,000 + &(km\,s$^{-1}$)\\
\noalign{\vskip 0.2mm}
\hline
\noalign{\vskip 0.2mm}
53451.2048&$-$12.9\\
53452.2318&$-$6.1\\
53453.2541& 5.3\\
53454.2702&$-$29.7\\
53455.2032&$-$32.7\\
\noalign{\vskip 0.2mm}
\hline
\end{tabular}
\label{tab-aysgr-data}
\end{table}

Our new RV data are listed in Table~\ref{tab-aysgr-data}.
The first epoch RV phase curve of AY~Sgr by \citet{J37} 
has been followed by our one 25\,000 days later. 
The merged RV phase curve in Fig.~\ref{fig-aysgr-vrad} 
shows a significant increase
in the $\gamma$-velocity during eight decades:
at JD\,2\,427\,640 $v_{\gamma} = -22.4$ km\,s$^{-1}$, while
at JD\,2\,453\,550 $v_{\gamma} = -15.5$ km\,s$^{-1}$
indicating the membership of AY~Sgr in a spectroscopic
binary system.
During the two observing seasons covered by our spectroscopic
observations, no shift in the $\gamma$-velocity is apparent.
Nevertheless, the larger amplitude of the phase curve based on 
Joy's data may be the consequence of the orbital motion in the
binary system during his five-year-long observational interval.

\subsection{ST~Velorum}
\label{stvel}

\paragraph*{Accurate value of the pulsation period}
\label{stvel-period}

\begin{figure}
\includegraphics[height=44mm, angle=0]{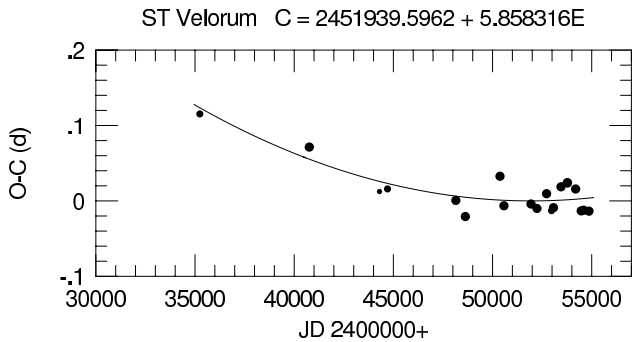}
\caption{$O-C$ diagram of ST~Vel. The plot can be
approximated by a parabola indicating a continuous period
increase}
\label{fig-stvel-oc}
\end{figure}

The brightness variability of ST~Vel (CD~$-$50$^{\circ}$~3533), 
$\langle V \rangle = 9.73$\,mag.) was suspected by Kapteyn
\citep{GI03} and it was reported as a new variable by 
\citet{Cetal09}. Being a Cepheid that pulsates in the 
fundamental mode, the $O-C$ diagram in Fig.\ref{fig-stvel-oc} 
has been constructed for the moments of brightness maxima listed 
in Table~\ref{tab-stvel-oc}. The final ephemeris obtained by a 
weighted parabolic least squares fit is:

\vspace{-1mm}
\begin{equation}
C = 2\,451\,939.5962 + 5.858\,316{\times}E 
\label{stvel-ephemeris}
\end{equation}
\vspace{-3mm}
$\phantom{mmmmm}\pm0.0025\phantom{}\pm 0.000\,006$

\noindent The second order term omitted from Eq.~\ref{stvel-ephemeris}
results in a continuous increase in the pulsation period
amounting to $(3.04 \pm 0.58)\times10^{-8}$ d/cycle, i.e., 16.36
s/century.

\begin{table}
\caption{$O-C$ values of ST~Velorum (description of the columns
is given in Sect.~\ref{ghcar-period} but the first column contains
the moments of brightness maxima)}
\begin{tabular}{l@{\hskip2mm}r@{\hskip2mm}r@{\hskip2mm}c@{\hskip2mm}l}
\hline
\noalign{\vskip 0.2mm}
JD$_{\odot}$ & $E\ $ & $O-C$ & $W$ & Data source\\
2\,400\,000 + &&&\\
\noalign{\vskip 0.2mm}
\hline
\noalign{\vskip 0.2mm}
35243.5109 &$-$2850 &+0.1153 &2& \citet{Wetal58} +\\
&&&& \citet{I61}\\
40762.0007 &$-$1908 &+0.0714 &3& \citet{P76}\\
44300.3645 &$-$1304 &+0.0124 &1& \citet{B08}\\
44704.5918 &$-$1235 &+0.0159 &2& \citet{E85}\\
48149.2664 &$-$647 &+0.0007 &3& {\it Hipparcos} \citep{ESA97}\\
48629.6270 &$-$565&$-$0.0207 &3& {\it Hipparcos} \citep{ESA97}\\
50375.4585 &$-$267 &+0.0327 &3& \citet{B08}\\
50574.6022 &$-$233&$-$0.0064 &3& \citet{B08}\\
51939.5921 & 0&$-$0.0041 &3& ASAS \citep{P02}\\    
52238.3603 &51&$-$0.0100 &3& ASAS \citep{P02}\\
52724.6201 &134 &+0.0096 &3& ASAS \citep{P02}\\
52976.5052 &177&$-$0.0129 &2&  {\it INTEGRAL} OMC\\
53070.2422 &193&$-$0.0090 &3& ASAS \citep{P02}\\
53451.0604 &258 &+0.0187 &3& ASAS \citep{P02}\\
53767.4149 &312 &+0.0241 &3& {\it INTEGRAL} OMC\\
53779.1313 &314 &+0.0239 &3& ASAS \citep{P02}\\
54189.2053 &384 &+0.0158 &3& ASAS \citep{P02}\\
54470.3755 &432&$-$0.0132 &3& ASAS \citep{P02}\\
54599.2593 &454&$-$0.0124 &3& ASAS \citep{P02}\\
54862.8823 &499&$-$0.0136 &3& ASAS \citep{P02}\\
\noalign{\vskip 0.2mm}
\hline
\end{tabular}
\label{tab-stvel-oc}
\end{table}

\paragraph*{Binarity of ST~Vel}
\label{stvel-bin}

\begin{figure}
\includegraphics[height=60mm, angle=0]{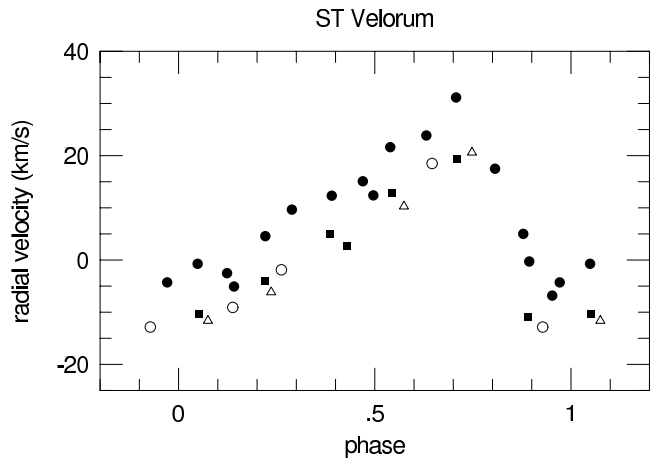}
\caption{Merged RV phase curve of ST~Vel. Triangles represent 
radial velocities obtained by \citet{Petal94},
our data are split into three parts: those from 2004 (empty 
circles), from 2005 (filled circles), and 2006 (filled
squares)}
\label{fig-stvel-vrad}
\end{figure}

\begin{figure}
\includegraphics[height=45mm, angle=0]{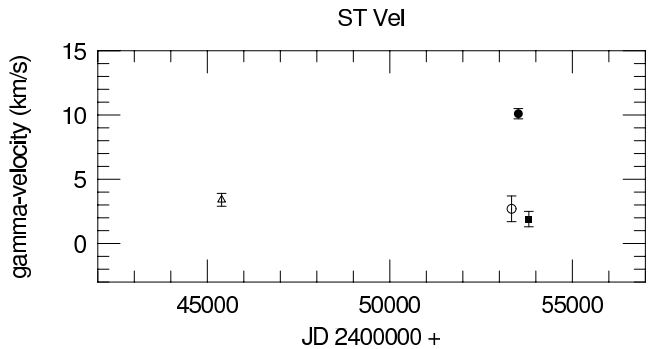}
\caption{Temporal variations in the $\gamma$-velocity of ST~Velorum.
The symbols are consistent with those used in 
Fig.~\ref{fig-stvel-vrad}}
\label{fig-stvel-vgamma}
\end{figure}

\begin{table}
\caption{New RV values of ST Velorum from the SSO spectra.
This is only a portion of the full version 
available online only}
\begin{tabular}{lr}
\hline
\noalign{\vskip 0.2mm}
JD$_{\odot}$ & $v_{\rm rad}$  \ \\
2\,400\,000 + &(km\,s$^{-1}$)\\
\noalign{\vskip 0.2mm}
\hline
\noalign{\vskip 0.2mm}
53310.2539&$-$12.9\\
53312.2108&$-$1.9\\
53364.2096&$-$9.1\\
53367.1854& 18.5\\
53450.9950&$-$6.8\\
\noalign{\vskip 0.2mm}
\hline
\end{tabular}
\label{tab-stvel-data}
\end{table}

\begin{table}
\caption{$\gamma$-velocities of ST~Velorum}
\begin{tabular}{lccl}
\hline
\noalign{\vskip 0.2mm}
Mid-JD & $v_{\gamma}$ & $\sigma$ & Data source \\
2\,400\,000+ & (km\,s$^{-1}$)& (km\,s$^{-1}$) & \\
\noalign{\vskip 0.2mm}
\hline
\noalign{\vskip 0.2mm}
45392 & 3.4 & 0.5 & \citet{Petal94}\\
53335 & 2.7 & 1.0 & present paper\\
53520 & 10.1& 0.4 & present paper\\
53808 & 1.9 & 0.6 & present paper\\
\noalign{\vskip 0.2mm}
\hline
\end{tabular}
\label{tab-stvel-vgamma}
\end{table}

The available RV data -- those by \citet{Petal94} and our
new ones listed in Table~\ref{tab-stvel-data} -- have been plotted 
in Fig.~\ref{fig-stvel-vrad}.
When folding the data into a phase curve, the period of 5.858316 day
given in the Eq.~\ref{stvel-ephemeris} was used but due to the
parabolic pattern of the $O-C$ graph, a tiny correction was applied
when plotting the data by \citet{Petal94}. It is noteworthy that our
own data show an excessive scatter that can be explained in terms of
the variation in the $\gamma$-velocity. This effect is clearly seen
in Fig.~\ref{fig-stvel-vgamma} where the annual $\gamma$-velocities
listed in Table~\ref{tab-stvel-vgamma} have been plotted. The pattern 
of the points in this figure implies that the orbital period can be
several hundred days, which is rather short among the spectroscopic
binaries containing a Cepheid component.

\section{Conclusions}
\label{concl}

We pointed out that six southern Galactic Cepheids,
GH~Carinae, V419~Centauri, V898~Centauri, AD~Puppis,
AY~Sagittarii, and ST~Velorum have a variable 
$\gamma$-velocity which implies their membership in 
spectroscopic binary systems. The available RV
data are insufficient to determine the orbital period and
other elements of the orbit. We can only state that the
orbital period of V419~Cen is several years, for V898~Cen
it is 2000-3000 days, for AD~Pup about 50 years, and for
ST~Vel several hundred days.

The value of the orbital period for spectroscopic binary 
systems involving a Cepheid component is often unknown: 
according to the on-line data base \citep{Sz03a}
the orbital period has been determined for about 20\% 
of the known SB Cepheids. Majority of the known orbital 
periods exceeds a thousand days.

Our finding confirms the previous statement by 
\citet{Sz03a} about the high percentage of binaries
among classical Cepheids and the observational selection 
effect hindering the discovery of new cases. 

Radial velocity data obtained prior to ours were instrumental
in discovering binarity of V419~Cen, AD~Pup, AY~Sgr, and
ST~Vel, while the spectroscopic binary nature of GH~Car and
V898~Cen has been discovered from our observations alone.

A companion star may have various effects on the photometric 
properties of the Cepheid component. Various pieces of 
evidence of duplicity based on the photometric criteria are
discussed by \citet{Sz03b} and \citet{KSz09}. As to our
targets, there is no obvious sign of companion from photometry
alone. This indicates that the companion star cannot be much 
hotter than the Cepheid component in either case. Nevertheless,
weak evidence of anomalous photometric behaviour was reported
for GH~Car by \citet{MF80} (abnormal phase shift between the 
light curves in different photometric bands) and for V419~Cen
by \citet{Oetal85} (anomalous behaviour when determining its
physical parameters with the CORS method). 
The strange spectral type of F3III for V898~Cen appearing in the
SIMBAD data base may be wrong because an F9 spectral type has
been deduced from our spectra. Further spectroscopic 
observations are necessary to characterise these binary systems.
In addition, accurate future photometric observations can be
instrumental in confirming the interpretation of the wavy pattern
superimposed on the parabolic $O-C$ graph
of AD~Pup in terms of a light-time effect.

Regular monitoring of the radial velocities of a large
number of Cepheids will be instrumental in finding 
more long-period spectroscopic binaries among Cepheids. 
Quite recently \citet{Evetal12} reported on their 
on-going survey for pointing out binarity of Cepheids 
from the existing RV data covering sufficiently long time 
interval. Radial velocity data to be obtained with the Gaia
astrometric space probe (expected launch: October 2013)
will certainly result in revealing new spectroscopic
binaries among Cepheids brighter than 13-14th magnitude
\citep{Eyetal12}.

When determining the physical properties (luminosity, 
temperature, radius, etc.) of individual Cepheids,
the effects of the companion on the observed parameters
(apparent brightness, colour indices, etc.) have to be
corrected for. This type of analysis, however, should be
preceded by revealing the binarity of the given Cepheid.

\section*{Acknowledgments} 

This project has been supported by the ESA PECS Project C98090, 
ESTEC Contract No.\,4000106398/12/NL/KML, the Hungarian OTKA 
Grants K76816, K83790, K104607, and MB08C 81013, as well as the 
European Community's Seventh Framework Program (FP7/2007-2013) 
under grant agreement no.\,269194,  and the ``Lend\"ulet-2009'' 
Young Researchers Program of the Hungarian Academy of Sciences. 
AD was supported by the Hungarian E\"otv\"os fellowship. 
AD has been supported by the J\'anos Bolyai Research Scholarship 
of the Hungarian Academy of Sciences. AD is very thankful 
to the staff at The Lodge in Siding Spring Observatory 
for their hospitality and the very nice food, making the 
time spent there lovely and special.
Part of the research leading to these results has received 
funding from the European Research Council under the European 
Community's Seventh Framework Programme (FP7/2007--2013)/ERC grant 
agreement n$^\circ$227224 (PROSPERITY).
The {\it INTEGRAL\/} photometric data, pre-processed by 
ISDC, have been retrieved from the OMC Archive at CAB (INTA-CSIC). 
Critical remarks by Dr. M\'aria Kun and the referee's suggestions
led to a considerable improvement in the presentation of the results.

\vspace*{1cm}
\noindent{\bf SUPPORTING INFORMATION}

\medskip
\noindent Additional Supporting Information may be found in the online version
of this article:

\medskip
\noindent{\bf Table 3.} New RV values of GH Carinae from the SSO spectra.\\
{\bf Table 4.} New CORALIE velocities of GH Carinae.\\
{\bf Table 6.} New RV values of V419 Centauri from the SSO spectra.\\
{\bf Table 9.} New RV values of V898 Centauri from the SSO spectra.\\
{\bf Table 13.} New RV values of AD Puppis from the SSO spectra.\\
{\bf Table 15.} New RV values of AY Sagittarii from the SSO spectra.\\
{\bf Table 17.} New RV values of ST Velorum from the SSO spectra.\\
(http://www.mnras.oxfordjournals.org/lookup/suppl/doi:10.1093/
mnras/stt027/-/DC1).

\medskip
\noindent Please note: Oxford University Press are not responsible for the
content or functionality of any supporting materials supplied by
the authors. Any queries (other than missing material) should be
directed to the corresponding author for the article.

\bsp

\label{lastpage}

\end{document}